\documentclass[aps,pra,twocolumn,superscriptaddress]{revtex4}
\usepackage{dcolumn}
\usepackage{graphicx}
\usepackage{amsmath}
\usepackage{amsfonts}
\usepackage{amssymb}
\usepackage{psfrag}
\usepackage{wrapfig}
\usepackage{subfigure}
\usepackage{makeidx}
\usepackage{bm}
\usepackage{epsf}
\usepackage{multirow}
\usepackage{upgreek}
\usepackage[normalem]{ulem}
\usepackage{amssymb}
\usepackage{mathrsfs}
\usepackage{xcolor}
\usepackage[colorlinks,urlcolor=blue,citecolor=blue]{hyperref}
\DeclareMathOperator{\sech}{sech}

\begin{document}
\title{Superheavy dark-bright soliton as a signature of  spatial  symmetry breaking transition in harmonically trapped Bose mixtures}
\date{\today}
	
\author{Zheng Gao}
\affiliation{School of Physics, Northwest University, Xi'an, 710127, China}
\author{Ling-Zheng Meng}
\affiliation{School of Science, Xi'an Technological University, Xi'an 710021, China}	\author{Jie Liu}
\affiliation{Graduate School, China Academy of Engineering Physics, Beijing 100193, China}
\author{Li-Chen Zhao}\email{zhaolichen3@nwu.edu.cn}
\affiliation{School of Physics, Northwest University, Xi'an, 710127, China}
\affiliation{NSFC-SPTP Peng Huanwu Center for Fundamental Theory, Xi'an 710127, China}
\affiliation{Shaanxi Key Laboratory for Theoretical Physics Frontiers, Xi'an 710127, China}
\affiliation{Fundamental Discipline Research Center for  Quantum Science and technology of Shaanxi Province}

\begin{abstract}
	We investigate the dynamics of a dark-bright soliton in harmonically trapped two-component Bose-Einstein condensates and reveal an interesting spontaneous spatial symmetry breaking driven by nonlinear interactions. When the interaction parameter crosses a threshold value, we find that the dark-bright soliton's motion demonstrates a transition from symmetric periodic oscillation about the origin to asymmetric oscillations offset from the origin. In particular, at the transition point, the effective soliton mass, determined by the ratio of inertial mass to physical mass, diverges. The underlying mechanism is uncovered by constructing trial wave functions and employing the Lagrangian variational method to obtain an effective potential in the quasiparticle picture, which changes from a single well to a double well. The anomalous ``superheavy soliton'' phenomenon is a direct consequence of the dark-bright soliton's physical mass vanishing at the transition point. We obtain the phase diagram of this spatial symmetry-breaking transition. Possible implications of our finding for quantum metrology are discussed.
\end{abstract}
\maketitle
	
\textit{Introduction} -- 
The dynamics of solitons driven by external fields has become one of the core research directions in soliton physics \cite{Kivshar1989RMP, Malomed2024Review, Kevrekidis2008Book, Frantzeskakis2010Review, Busch2000PRL}. It enables the characterization of multiple intrinsic properties of solitons, including inertial mass and wave structure, and further facilitates the exploration of their anomalous transport behaviors \cite{Busch2001PRL, Yefsah2013Nature, Meng2022NegativeMass, Khamehchi2017NegativeMass, Asad2013SolitonDiode}. Among them, the motion of solitons in a harmonic trap has been widely discussed, partly due to the fact that it is a common, easily prepared potential in quantum gases, and the oscillation frequency can directly reflect the intrinsic properties of solitons \cite{Busch2000PRL, Konotop2004PRL, Becker2008NaturePhys, Weller2008PRL, Ku2014PRL}. In a quasi-one-dimensional harmonically trapped Bose gas, the oscillation frequency of a dark soliton is $1/\sqrt{2}$ times the trap frequency \cite{Busch2000PRL}, while a bright soliton oscillates at the trap frequency. For a unitary Fermi gas, the dark soliton frequency is approximately $1/\sqrt{3}$ times the trap frequency \cite{Antezza2007PRA, Scott2011PRL}. The frequency modifications are induced by the strikingly large inertial mass of dark solitons. The oscillation frequency can be further decreased by coupling to a bright soliton \cite{Becker2008NaturePhys, Busch2001PRL,Guan2022CPB} or to atoms of Andreev bound states \cite{Antezza2007PRA}. One can thus consider the soliton becomes heavier and is harder to move. Heavy solitons have been found in Fermi gases, for which the soliton effective mass is more than 200 times their bare mass \cite{Yefsah2013Nature, heavy1}. This motivates us to look for heavy solitons in Bose gases and to check whether there are some striking differences between them.
	
Dark-bright (DB) solitons---a composite structure confining a bright soliton within a dark soliton \cite{Busch2001PRL, likebehavior1, solitondynamic3, solitondynamic4, fields3, solitondynamic7, solitondynamic1}---have been found to admit both positive- and negative-mass branches in Bose gases with hyperfine states, which cause solitons to undergo Josephson-like oscillations \cite{Negativemass2, jiangnan, Meng2022NegativeMass, likebehavior7, josephson}, and this remarkable phenomenon has been recently observed experimentally \cite{Rabec2025}. Recalling that the bright soliton and dark soliton components possess positive and negative effective mass, respectively, the relative dominance of the two inertial mass branches can be tuned by nonlinearities \cite{Negativemass2, Meng2022NegativeMass, Khamehchi2017NegativeMass}. Considering that the nonlinear interaction strengths and the mass ratio of the two atomic species provide much more freedom in Bose mixtures than in usual Bose gases with hyperfine states \cite{Chin2010RMP, Pilch2009PRA, solitondynamic8}, we expect that the DB soliton could be used to achieve the observation of a superheavy phenomenon in Bose mixtures.
	
In this paper, we investigate the dynamics of a dark-bright soliton in harmonically trapped Bose mixtures. The results indicate that the soliton motion exhibits different characteristics depending on the interaction strengths and the mass ratio of the two components. In particular, a striking one-sided oscillation is observed, which differs significantly from the usual oscillation around the center of the trapping potential. At the boundary separating these regimes, a superheavy phenomenon appears, where the soliton remains pinned at its initial position.
By formulating a generalized variational framework that intrinsically accounts for the non-uniformity of the Thomas-Fermi background, we derive a comprehensive effective energy landscape governing the soliton's motion, which can be viewed as an effective potential. We demonstrate that the superheavy state marks an exact threshold where this effective potential transitions from a symmetric single well to a degenerate double-well configuration, thereby inducing symmetry breaking of the soliton motion in the harmonic trap.
This striking behavior is physically analogous to the interaction-driven transition from tunneling to macroscopic nonlinear self-trapping in double-well potentials \cite{Fu2006, Josephsoneffect}, but the phase transition here is of first order. The high sensitivity to nonlinear interactions could establish a new paradigm for critical quantum metrology in many-body systems.
	
\textit{Superheavy phenomenon and spatial symmetry breaking transition of dark-bright soliton in a trap} -- 
We consider a two-component Bose mixture composed of $^{133}$Cs and $^{87}$Rb atoms, noting that the mixtures exhibit widely tunable Feshbach resonances \cite{Pilch2009PRA,McCarron2011PRA}. The system is confined in harmonic traps and elongated into a cigar shape. Within the mean-field approximation, the dynamics of this quasi-one-dimensional system is governed by the following dimensionless coupled Gross-Pitaevskii (GP) equations
\begin{subequations}\label{eq:GP_equations}
	\begin{align}
		i\partial_t\psi_d &= \left[ -\frac{1}{2m_d}\partial_x^2 + V_d(x) + \left(g_d|\psi_d|^2 + g_{bd}|\psi_b|^2\right) \right] \psi_d, \label{eq:GP_d} \\[1ex]
		i\partial_t\psi_b &= \left[ -\frac{1}{2}\partial_x^2 + V_b(x) + \left(g_{bd}|\psi_d|^2 + g_b|\psi_b|^2\right) \right] \psi_b. \label{eq:GP_b}
	\end{align}
\end{subequations}
where $\psi_d$ and $\psi_b$ denote the dark and bright soliton wave functions in the two components, respectively. The length, time, and energy are measured in units of $\ell_\perp = \sqrt{\hbar/(m_{{\rm Rb}}\omega_\perp)}$, $\omega_\perp^{-1}$, and $\hbar\omega_\perp$, respectively.
By scaling atom mass of the bright soliton component  to be unity, the the mass ratio is $m_d = m_{{\rm Cs}}/m_{{\rm Rb}} = 1.52$. The external harmonic trapping potentials are $V_d(x) = \frac{1}{2}m_d \omega^2 x^2$ and $V_b(x) = \frac{1}{2}\omega^2 x^2$ with a frequency $\omega$. The coefficients $g_{{\rm i}}=2a_{{\rm ii}}n_0$ and $g_{{\rm ij}}=2a_{{\rm ij}}n_0$ characterize the intra- and inter-species interaction strengths, where $n_0$ denotes one-dimensional density.
	
To investigate the dynamics of DB solitons in the trap, we can perform the Lagrangian variational method to derive the DB solution as a starting point (see Appendix) while ignoring the external traps \cite{Kivshar1989RMP, exactsolution}, provided that the trap frequency $\omega \ll 1$. The weak external traps ensure that the background density variation across the soliton scale is much smaller than the soliton amplitude. Then, taking the external traps into account, the wave functions for the DB soliton can be written as $\psi_d = \sqrt{\rho_{{\rm TF}}} \{i\sqrt{1 - f_d^2} + f_d \tanh[w_d(x - x_c)]\}$ and $\psi_b = f_b \sech[w_b(x - x_c)] e^{i[\xi + \phi (x - x_c)]}$, where $\rho_{{\rm TF}}=\max [ 1 - \frac{m_d\omega^2x^2}{2g_d},0]$ is the background density for the dark soliton excitation under the Thomas-Fermi (TF) approximation \cite{fields5, Baym1996PRL, Stringari1996PRL}. The other parameters denote the intrinsic properties of solitons. $x_c$ denotes the center position of the soliton complex, $f_d$ ($f_b$) and $w_d$ ($w_b$) denote the amplitude and width of dark (bright) soliton. $\xi$ and $\phi$ denote phase and wave vector of the bright soliton component.
\begin{figure}[t]
	\centering
	\includegraphics[width=\columnwidth]{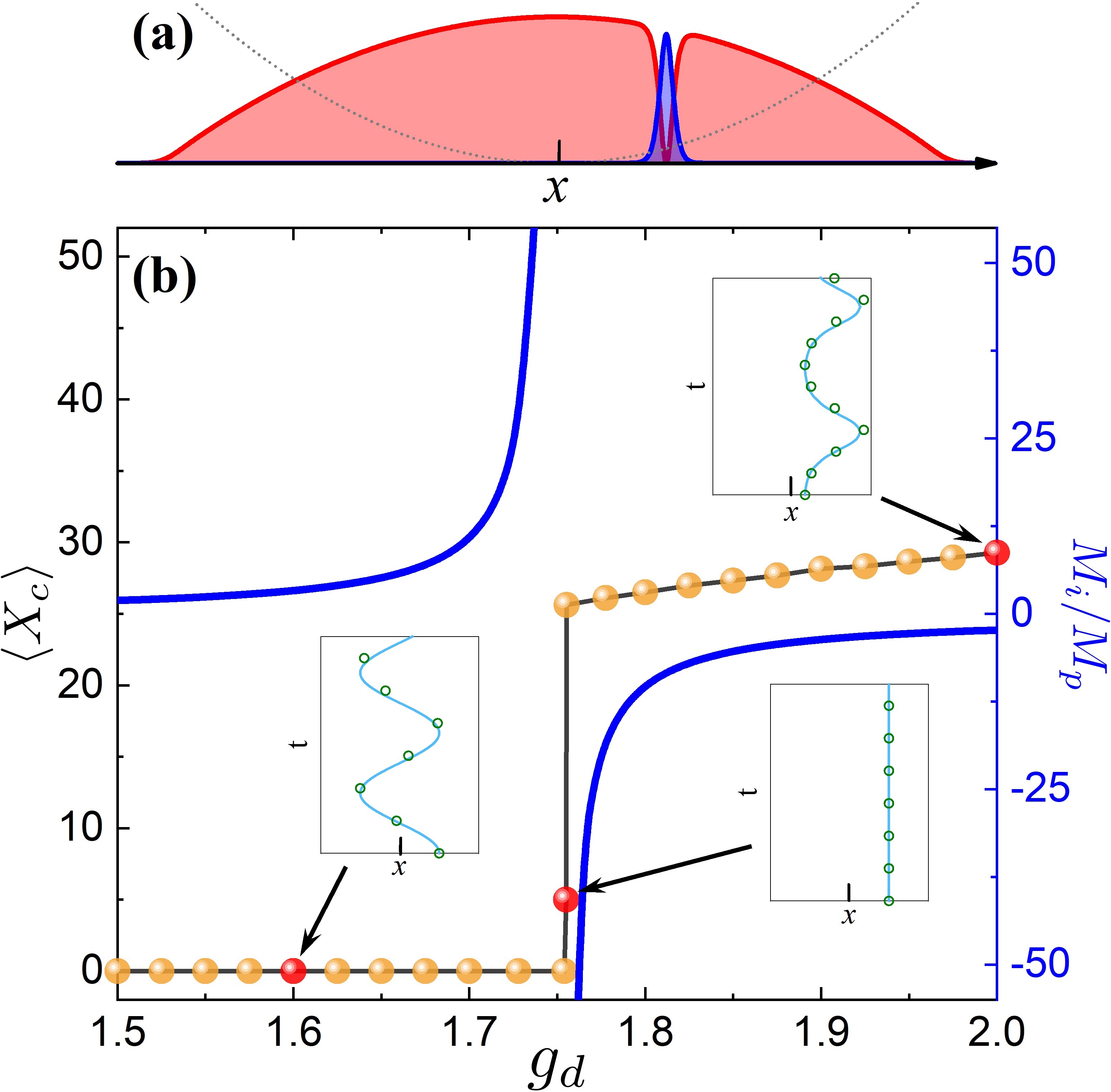}
	\caption{Dynamics of a dark-bright soliton in a harmonic trap. (a) Schematic representation of the dark-bright soliton in the harmonic potential (dotted line). (b) Time-averaged position of the soliton $\langle X_c \rangle$ (orange spheres) and soliton effective mass $M_i/M_p$ (blue solid line) as functions of $g_d$. The soliton effective mass diverges and the superheavy phenomenon occurs at the threshold, indicating a phase transition between the two distinct dynamical regimes. The three insets show the real-time trajectories of the soliton at representative values $g_d = 1.6$, $1.755$, and $2$ (marked by red spheres). In these insets, solid lines correspond to the analytical predictions from quasiparticle theory, and open circles are the numerical simulation results. The parameters are $g_b = 0.3$, $g_{bd} = 1$, $N_b = 4.7476$, $\omega = 0.01$, and the initial position $x_{c0} = 5$.}
	\label{fig:evolution}
\end{figure}
	
We study the motion of a DB soliton in a harmonic trap by numerically simulating the coupled GP equations with the soliton initially placed at $x_{c0} = 5$ (the schematic representation of the initial setting is shown in Fig.~\ref{fig:evolution}(a)). In this paper, we fix $g_{b}$ and $g_{bd}$, and only change the intra-component interaction of the dark soliton component $g_{d}$ without loss of generality. We observe that the DB soliton exhibits three distinct dynamical regimes when the nonlinear interaction strength changes. As an example, we characterize the soliton motion by the average center position during one period $\langle X_c \rangle$ using the parameters $N_b = 4.7476$, $\omega = 0.01$, $g_b = 0.3$, $g_{bd} = 1$ (the dimensionless bright-component norm $N_b=\int |\psi_b|^2{\rm d}x$ corresponds to the atom number $\mathcal{N}_b=n_0\ell_\perp N_b\simeq1.3\times10^2$ atoms for $^{87}$Rb with $a_{bb}\simeq100a_0$ and $\ell_\perp=1.0\mu{\rm m}$).
The numerical results are shown as orange spheres in Fig.~\ref{fig:evolution}(b). For $g_d < 1.755$, the soliton exhibits normal oscillation around the center of the harmonic trap (as shown in the left inset), and thus $\langle X_c \rangle = 0$. The oscillation frequency is much lower than the trap frequency and even lower than that of scalar dark solitons \cite{Busch2000PRL,Busch2001PRL,Becker2008NaturePhys}. A striking dynamical behavior emerges at the critical value $g_d = 1.755$. The periodic motion completely ceases, and the soliton becomes permanently pinned at its initial off-center position with $\langle X_c \rangle = 5$. The phenomenon can be termed superheavy because the oscillation period diverges at this point, in comparison with the heavy ones reported in Fermi gases \cite{Yefsah2013Nature, heavy1} for which the soliton effective mass is more than 200 times its bare mass. For $g_d > 1.755$, the motion of the soliton demonstrates a striking spatial symmetry breaking character in the symmetric harmonic trap, and it admits one-sided oscillation, resulting in a substantial non-zero $\langle X_c \rangle$. One example with $g_d = 2$ is shown in the upper inset of Fig.~\ref{fig:evolution}(b). This transition indicates that the system undergoes a first-order phase transition, in sharp contrast to the second-order transitions for the transition from Josephson oscillation to self-trapping with varying nonlinear parameter in double-well potentials \cite{Smerzi1997PRL, Albiez2005PRL, Fu2006}. The transition from normal oscillation to one-sided oscillation was also demonstrated by the ferrodark soliton dynamics in spin-1 superfluids \cite{jiangnan}, where the soliton position plays an essential role in the transition.

\textit{Quasiparticle theory for superheavy phenomenon and spatial symmetry breaking behaviors} -- 
To analytically describe and fully characterize the three distinct dynamical behaviors, we derive the classical kinetic equations for the soliton within a quasiparticle picture based on energy conservation. The total energy $E_{tot}=E_s + E_p$ is decomposed into the excitation energy of the solitons $E_s$ and the external potential energy $E_p$. Accounting for the spatial modulation of the inhomogeneous harmonic trap, we obtain the excitation energy functional $E_s = \int [ \frac{1}{2m_d}|\partial_x\psi_d|^2 + \frac{1}{2}|\partial_x\psi_b|^2 + \frac{g_b}{2}|\psi_b|^4 + \frac{g_d}{2}(|\psi_d|^2 - \rho_{\text{TF}})^2 + g_{bd}|\psi_b|^2(|\psi_d|^2 - \rho_{\text{TF}})]{\rm d}x$ \cite{exactsolution,Kivshar1995OC}. Meanwhile, the external trapping energy is given by $E_p = \int [ \frac{1}{2}m_d\omega^2x^2|\psi_d|^2 + \frac{1}{2}\omega^2x^2|\psi_b|^2 ] {\rm d}x$.

For simplicity and to capture the essential physics, we first consider the case where the soliton is initially positioned close to the trap center and its velocity is sufficiently low ($v = \dot{x}_c \simeq 0$). In this case, $(x_c/R_{{\rm TF}})^2 \ll 1$, where $R_{{\rm TF}} = \sqrt{2g_d/(m_d\omega^2)}$ is the Thomas-Fermi radius, which renders the core density gradient negligible, thus validating the uniform-background approximation $\rho_{{\rm TF}} \approx 1$.
By substituting the wave functions of the two components, one can obtain the excitation energy as $E_s(v) = E_{s,0} + \frac{1}{2}M_i v^2$, which is an even function of the velocity. The rest energy $E_{s,0} = E_s(0)$ and the inertial mass $M_i = 2 \partial E_s/\partial(v^2)|_{v=0}$ are evaluated using the coupled equations, Eqs.~\eqref{eq:reduced_dynamics_wb} and \eqref{eq:reduced_dynamics_wd} of the Appendix.
Correspondingly, the external potential energy is $E_p = E_{p,0} + \frac{1}{2}M_p\omega^2x_c^2$, where $E_{p,0}$ is the static energy offset and $M_p = \int |\psi_b|^2 + m_d(|\psi_d|^2 - \rho_{{\rm TF}})dx= N_b - 2m_df_d^2/w_d$ defines the physical mass. The two masses of the DB soliton change with the nonlinear parameters. Then the energy conservation law $\frac{{\rm d}}{{\rm d}t}(\frac{1}{2}M_i\dot{x}_c^2 + \frac{1}{2}M_p\omega^2x_c^2)= 0$ yields the kinetic equation for the soliton,
\begin{equation}
	\frac{M_i}{M_p}\ddot{x}_c + \omega^2 x_c = 0.
	\label{eq:equations 2}
\end{equation}
This equation describes the motion of the soliton, with effective mass $M_i/M_p$, in a harmonic trap with frequency $\omega$. Moreover, the motion of a DB soliton can also be regarded as that of a quasiparticle in an effective potential $U_{\rm eff}(x_c) = \frac{1}{2}\omega_{\rm eff}^2x_c^2$, with the effective oscillation frequency defined as $\omega_{\rm eff} = \sqrt{M_p/M_i}\omega$ from the above equation.
	
\begin{figure}[t]
	\centering
	\includegraphics[width=\columnwidth]{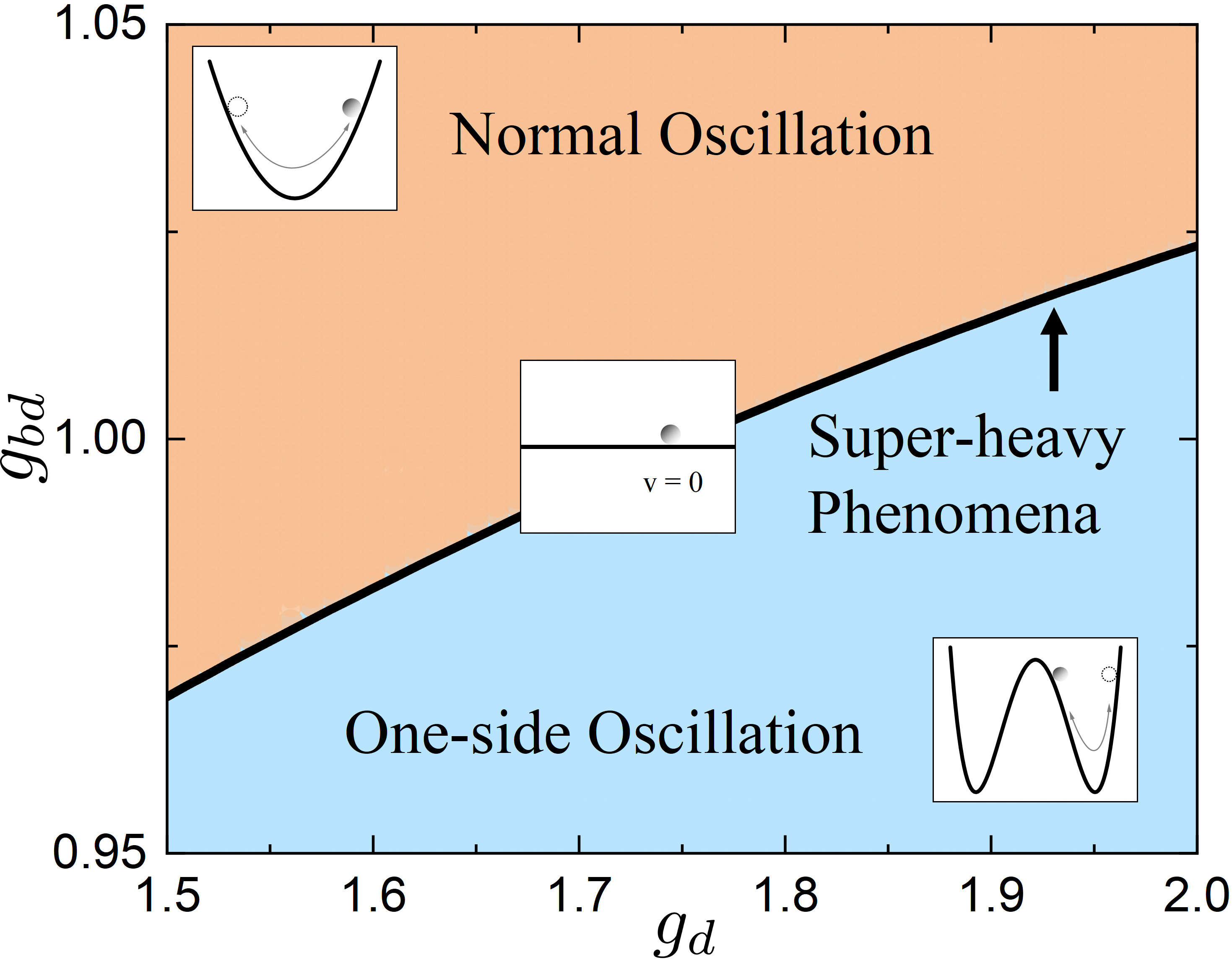}
	\caption{Dynamical phase diagram of the dark-bright soliton in the $g_d$-$g_{bd}$ parameter space. The orange and blue regions denote the normal and one-sided oscillation phases, respectively. The black solid line marks the critical phase boundary associated with the superheavy phenomenon, where the soliton becomes completely stationary ($v=0$) and which separates the two oscillatory regimes. The three insets illustrate the underlying effective potentials $U_{\rm eff}$ governing each dynamical regime: a single-well potential for normal oscillation, a flat potential landscape for the superheavy state, and a double-well potential for one-sided oscillation. The parameters are $g_b = 0.3$, $N_b = 4.7476$, $\omega = 0.01$, and initial position $x_{c0} = 5$.}
	\label{fig:potential_phase_diagram}
\end{figure}

The inertial mass $M_i$ is intrinsically negative, and the physical mass $M_p$ is also negative when $g_d$ is small, yielding a real effective frequency $\omega_{\rm eff}^2 > 0$ (the positive soliton effective mass $M_i/M_p$ as the left branch in Fig.~\ref{fig:evolution}(b)). Therefore, the DB soliton can exhibit normal oscillation like a classical particle with fixed mass.
This dynamically stable regime corresponds precisely to the orange region of the phase diagram in Fig.~\ref{fig:potential_phase_diagram}.
This quasiparticle formulation naturally parallels the seminal theoretical treatments of scalar dark solitons in weakly inhomogeneous backgrounds \cite{Busch2001PRL,Konotop2004PRL}, but the oscillation frequency is determined by the subtle interplay between $M_p$ and $M_i$. The analytical trajectory derived from Eq.~\eqref{eq:equations 2} exhibits excellent quantitative agreement with numerical simulations, as shown in the left inset of Fig.~\ref{fig:evolution}(b).
	
When the nonlinear interaction strength $g_d$ increases, the corresponding physical mass $M_p$ increases and approaches zero, leading to a decrease in the effective frequency $\omega_{\rm eff}$. Intuitively, once $M_p=0$, the soliton effective mass diverges. Consequently, $\omega_{\rm eff}$ becomes zero and the oscillation period diverges. The DB soliton becomes unmovable, which accounts for the ``superheavy'' phenomenon. From the perspective of a quasiparticle in an effective potential, it is static because $U_{\rm eff}=0$.
This superheavy regime corresponds precisely to the black line in Fig.~\ref{fig:potential_phase_diagram}. The superheavy phenomenon in our bosonic DB complex stems strictly from this macroscopic cancellation of physical mass, in contrast to the ``heavy solitons'' in fermionic superfluids \cite{Yefsah2013Nature,heavy1}, which originate from microscopic quantum fluctuations and Andreev bound state filling. It should be noted that the superheavy phenomenon has also been reported in a spin-1 Bose condensate, in which the soliton cannot be driven by the external force due to the conservation of magnetization when the quadratic Zeeman energy tends to be zero \cite{Yu2022PRL}.
	
Crucially, as $g_d$ increases further beyond the critical threshold of the superheavy regime, $M_p$ tends to become positive and a negative soliton effective mass $M_i/M_p$ (right branch in Fig.~\ref{fig:evolution}(b)) can emerge, rendering the above effective harmonic trap picture invalid ($\omega_{\rm eff}^2 < 0$).
Physically, this transforms the center of $U_{\rm eff}$ from a stable minimum into an unstable local maximum, propelling the soliton away from the trap center.
This qualitatively explains the emergence of the above first-order phase transition. The inherent uniform-background approximation fails to describe the ensuing large-amplitude, off-center trajectories, since the theoretical treatment bifurcates according to the background inhomogeneity. We try to give a generalized theoretical treatment that explicitly incorporates the inhomogeneous Thomas-Fermi density profile, for exploring this symmetry-broken regime.
	
We derive more precise forms of the excitation energy and potential energy by using the original form of $\rho_{{\rm TF}}$ and obtain the full functional expressions $E_s = E_{s,0} + C_{s,2} x_c^2 + C_{s,4} x_c^4$ and $E_p = E_{p,0} + C_{p,2} x_c^2 + C_{p,4} x_c^4$ (see details in the Appendix). The generalized equation of motion is obtained as
\begin{equation}\label{GE}
	2 \left[ \frac{\partial E_{s,0}}{\partial (v^2)} + \frac{\partial \tilde{E}_p}{\partial (v^2)} \right] \ddot{x}_c + \frac{\partial \tilde{E}_p}{\partial x_c} = 0,
\end{equation}
where the modified potential energy $\tilde{E}_p(v^2, x_c)=E_{p,0}+ (C_{s,2}+ C_{p,2}) x_c^2 + (C_{s,4}+ C_{p,4}) x_c^4$. This $\tilde{E}_p$ also generally depends on the moving velocity because the soliton profiles depend on the velocity.
The term $2 \frac{\partial \tilde{E}_p}{\partial (v^2)}$ clearly indicates that the mass of the soliton generally depends on its position in the harmonic trap. This explicitly links the acceleration to the gradient of a generalized effective potential $U_{\rm eff} = \int_0^{x_c} \frac{2v(\zeta)\zeta[(C_{s,2}+ C_{p,2})+2(C_{s,4}+ C_{p,4}) \zeta^2]}{\partial_v(E_{s,0} + E_{p,0})+\partial_v(C_{s,2}+ C_{p,2})\zeta^2+\partial_v (C_{s,4}+ C_{p,4}) \zeta^4} d\zeta + {\rm const.}$, which can be calculated according to Eqs.~\eqref{eq:lda_constraint_fb}-\eqref{eq:lda_constraint_wd} of the Appendix.
This reveals a profound structural transition. As the interaction strength $g_{bd}$ or $g_d$ crosses a critical threshold, $U_{\rm eff}$ bifurcates from a single harmonic well into a double well, marking the entry into the blue region of the phase diagram in Fig.~\ref{fig:potential_phase_diagram}. This double-well framework perfectly reproduces the one-sided oscillation dynamics (e.g., $g_d = 2$ in the right inset of Fig.~\ref{fig:evolution}(b)).
	
If we mainly consider soliton motion with low speed, the soliton profile is nearly invariant, which makes $\partial E_p/\partial (v^2) \approx 0$. Then Eq.~\eqref{GE} can be simplified as
\begin{equation}
	M_i \ddot{x}_c +   2 (C_{s,2}+ C_{p,2}) x_c + 4 (C_{s,4}+ C_{p,4})  x_c^3 = 0.
\end{equation}
The coefficients $C_{s,2},C_{p,2},C_{s,4}$, and $C_{p,4}$ can be approximately calculated by using Eqs.~\eqref{eq:lda_constraint_fb}-\eqref{eq:lda_constraint_wd} of the Appendix with $v=0$. Furthermore, it is safe to neglect the higher-order term for this small-amplitude dynamics localized near the center of the harmonic trap, and then the above equation reduces to Eq.~\eqref{eq:equations 2}.
	
Our framework reveals the macroscopic symmetry breaking as a universal topological feature driven purely by intra- and inter-component nonlinearities, without relying on any spin exchanges \cite{jiangnan}. By generalizing the foundational paradigms of harmonic center-of-mass motion \cite{Busch2000PRL, Busch2001PRL, Konotop2004PRL, Becker2008NaturePhys}, our velocity-dependent effective potential establishes a rigorous mathematical bridge connecting the traditional oscillation regime directly to this extreme symmetry-broken phase. Nevertheless, there are some obvious deviations between the analytical trajectories and numerical simulations when the soliton approaches the condensate edge.
	
\textit{Conclusion} -- In summary, we have identified an anomalous ``superheavy'' phenomenon in the dynamics of dark-bright solitons within harmonically trapped Bose mixtures.
Characterized by a dramatic divergence of the oscillation period, this superheavy state is revealed as the macroscopic signature of a symmetry-breaking phase transition from normal to one-sided oscillations.
By developing a generalized variational framework that rigorously incorporates the non-uniformity of the background density, we obtain a complete phase diagram for the soliton dynamics, which includes normal oscillations, the superheavy phenomenon, and one-sided oscillations.
These three distinct dynamical phases correspond to effective potentials that exhibit symmetric single wells, flat landscapes, and double-well configurations, respectively.
Our results establish this macroscopic potential reshaping as a consequence of varying multi-component nonlinearities. The first-order phase transition induced by nonlinear parameters could be used to design ultrasensitive quantum metrology devices.

\section*{ACKNOWLEDGMENTS}
L.-C. Zhao was supported by the National Natural Science Foundation of China (Contracts No. 12375005, No. 12235007, and No. 12247103).
L.-Z. Meng was supported by the National Natural Science Foundation of China (Contract No. 12405002), the Young Talent Fund of Association for Science and Technology in Shaanxi, China (Grant No. 20250516), and the Natural Science Foundation of Shaanxi Provincial Department of Education (Grant No. 24JK0490).
	
\begin{widetext}
	\begin{appendix}
		\section{Derivation of dark-bright soliton solutions by means of the variational method}
		We obtain exact dark-bright soliton solutions of a two-component system by employing the variational method. Specifically, we introduce the Lagrangian
		\begin{align}
			L = \int \Bigg[ &\frac{i}{2} (\psi_d^* \partial_t \psi_d - \psi_d \partial_t \psi_d^*) \left( 1 - \frac{1}{|\psi_d|^2} \right) - \frac{|\partial_x \psi_d|^2}{2 m_d} - \frac{g_d}{2} (|\psi_d|^2 - 1)^2 \notag \\
			& + \frac{i}{2} (\psi_b^* \partial_t \psi_b - \psi_b \partial_t \psi_b^*) - \frac{1}{2} |\partial_x \psi_b|^2 - \frac{g_b}{2} |\psi_b|^4 - g_{bd} |\psi_b|^2 (|\psi_d|^2 - 1) \Bigg] \, {\rm d}x. \label{eq:lagrangian}
		\end{align}
		The terms $1 - \frac{1}{|\psi_d|^2}$ and $|\psi_d|^2 - 1$ were introduced in the Lagrangian density for a dark soliton mainly because a dark soliton is a density dip with a striking phase jump, in sharp contrast to a bright soliton \cite{Kivshar1995OC}. In the variational method, we assume the following trial wave functions for the two components
		\begin{align}
			\psi_d &= i \sqrt{1 - f_d^2} + f_d \tanh[w_d (x - x_c)]\vphantom{\Bigg|}, \label{eq:psi_d} \\
			\psi_b &= f_b \mathrm{sech}[w_b (x - x_c)] e^{i [\xi + (x - x_c) \phi]}. \label{eq:psi_b}
		\end{align}
		The parameter $f_{b,d}$ and $w_{b,d}$ respectively describe the amplitude and width of the bright and dark soliton. The central position of the soliton is $x_c$. The time-dependent phase of the bright soliton is $\xi$, and $\phi$ is related to its velocity. Substituting Eqs.~\eqref{eq:psi_d} and \eqref{eq:psi_b} into Eq.~\eqref{eq:lagrangian} and integrating over space from $-\infty$ to $\infty$ yields the effective Lagrangian
		\begin{align}
			L ={}& 2 \left( \arcsin f_d - f_d \sqrt{1 - f_d^2} \right) \dot{x}_c + \frac{2 f_b^2}{w_b} (\phi \dot{x}_c - \dot{\xi}) - \frac{2 f_d^2 w_d}{3 m_d} - \frac{2 g_d f_d^4}{3 w_d} \notag \\
			& - \frac{f_b^2}{3 w_b} \left( w_b^2 + 3\phi^2 + 2 g_b f_b^2 \right) + g_{bd} f_b^2 f_d^2 G. \label{eq:effective_lagrangian}
		\end{align}
		Here, the spatial overlapping integral and its partial derivatives are defined as $G = \int_{-\infty}^{\infty} \mathrm{sech}^2(w_b x) \mathrm{sech}^2(w_d x) \, {\rm d}x$, $\partial_{w_b} G = - \int_{-\infty}^{\infty} 2x \mathrm{sech}^2(w_b x) \mathrm{sech}^2(w_d x) \tanh(w_b x) \, {\rm d}x$, and $\partial_{w_d} G = - \int_{-\infty}^{\infty} 2x \mathrm{sech}^2(w_b x) \mathrm{sech}^2(w_d x) \tanh(w_d x) \, {\rm d}x$. By applying the Euler-Lagrange equations, $\frac{{\rm d}}{{\rm d}t} \left[ \frac{\partial L}{\partial \dot{\alpha}} \right] = \frac{\partial L}{\partial \alpha}$, with respect to the set of variational parameters $\alpha \in \{f_{d,b}, w_{d,b}, \xi, \phi, x_c\}$, we derive the following set of dynamical equations,
		\begin{subequations}\label{eq:evolution_equations}
			\begin{align}
				\makebox[5.0em][l]{$\alpha = f_b$} &:\quad 4 g_b f_b^2 + w_b^2 + 3 \phi^2 - 6 (\phi \dot{x}_c - \dot{\xi}) - 3 g_{bd} f_d^2 w_b G = 0 \vphantom{\Bigg|}, \label{eq:euler_fb} \\
				\makebox[5.0em][l]{$\alpha = w_b$} &:\quad 2 g_b f_b^2 - w_b^2 + 3 \phi^2 - 6 (\phi \dot{x}_c - \dot{\xi}) + 3 g_{bd} f_d^2 w_b^2 \partial_{w_b} G = 0 \vphantom{\Bigg|}, \label{eq:euler_wb} \\
				\makebox[5.0em][l]{$\alpha = f_d$} &:\quad 2 w_d^2 + 4 m_d g_d f_d^2 - \frac{6 m_d f_d w_d \dot{x}_c}{\sqrt{1 - f_d^2}} - 3 m_d g_{bd} f_b^2 w_d G = 0 \vphantom{\Bigg|}, \label{eq:euler_fd} \\
				\makebox[5.0em][l]{$\alpha = w_d$} &:\quad 2 w_d^2 - 2 m_d g_d f_d^2 - 3 m_d g_{bd} f_b^2 w_d^2 \partial_{w_d} G = 0 \vphantom{\Bigg|}, \label{eq:euler_wd} \\
				\makebox[5.0em][l]{$\alpha = \phi$} &:\quad \frac{2 f_b^2}{w_b}(\dot{x}_c - \phi) = 0 \vphantom{\Bigg|}, \label{eq:euler_phi} \\
				\makebox[5.0em][l]{$\alpha = x_c$} &:\quad \frac{{\rm d}}{{\rm d}t} \left[ \frac{2 f_b^2 \phi}{w_b} + 2 \left( \arcsin f_d - f_d \sqrt{1 - f_d^2} \right) \right] = 0 \vphantom{\Bigg|}, \label{eq:euler_xc} \\
				\makebox[5.0em][l]{$\alpha = \xi$} &:\quad \frac{{\rm d}}{{\rm d}t} \left[ \frac{2 f_b^2}{w_b} \right] = 0 \vphantom{\Bigg|}. \label{eq:euler_xi}
			\end{align}
		\end{subequations}
		Notice that the above coupled equations lead to the relations $\dot{x}_c = \phi$ and $N_b = 2f_b^2/w_b \equiv {\rm constant}$. Substituting these back greatly simplifies the remaining two coupled equations:
		\begin{align}
			4 m_d g_d (g_b N_b + 2 w_b) - 3 g_{bd} w_d^2 ( 4 - 3 m_d g_{bd} N_b w_b \partial_{w_d} G ) ( G + w_b \partial_{w_b} G ) &= 0, \label{eq:reduced_dynamics_wb} \\[2ex]
			\frac{4 w_d}{m_d} - g_{bd} N_b w_b G - 2 g_{bd} N_b w_b w_d \partial_{w_d} G - 4 \dot{x}_c w_d \sqrt{\frac{4 - 3 m_d g_{bd} N_b w_b \partial_{w_d} G}{4 g_d m_d - w_d^2 (4 - 3 m_d g_{bd} N_b w_b \partial_{w_d} G)}} &= 0. \label{eq:reduced_dynamics_wd}
		\end{align}
		Specifying the physical parameters ($N_b$, $m_d$, $\dot{x}_c$, and interaction strengths) reduces Eqs.~\eqref{eq:reduced_dynamics_wb} and \eqref{eq:reduced_dynamics_wd} to a solvable nonlinear system for $w_b$ and $w_d$. The numerically obtained soliton profiles then enable the direct evaluation of macroscopic properties.
		
		\section{The derivations for the general kinetic equation for soliton motion}
		The dynamics of the system are governed by the total energy functional:
		\begin{align}
			E_{\text{tot}} = \int_{-R_{\rm TF}}^{R_{\rm TF}} {\rm d}x \Bigg(
			& \frac{|\partial_x \psi_d|^2}{2m_d}+ \frac{1}{2} |\partial_x \psi_b|^2+ \frac{g_b}{2} |\psi_b|^4 + \frac{g_d}{2} \left[ |\psi_d|^2 - \left(1 -(x/R_{\rm TF})^2\right) \right]^2 \notag \\
			& + g_{bd} |\psi_b|^2 \left\{ |\psi_d|^2 - [1 - (x/R_{\rm TF})^2] \right\} + \frac{1}{2} m_d \omega^2 x^2 |\psi_d|^2+ \frac{1}{2} \omega^2 x^2 |\psi_b|^2\Bigg).
		\end{align}
		To analytically investigate the dynamics of the coupled system, we modify the trial ansatz $\psi_d = \{i \sqrt{1 - f_d^2} + f_d \tanh[w_d (x - x_c)] \} \sqrt{1 - (x/R_{\rm TF})^2}$ for the dark soliton, and $\psi_b = f_b \operatorname{sech}[w_b (x - x_c)] \exp\{ i [\xi + (x - x_c) \phi] \}$ for the bright soliton. To calculate the dark soliton kinetic energy $E_{kd}$, we introduce the relative coordinate $x_{th} = x - x_c$. Substituting the trial wave function and integrating the exactly solvable terms, we obtain the kinetic energy of dark soliton component as
		\begin{align}
			E_{kd} &= \int_{-R_{\rm TF}}^{R_{\rm TF}} \frac{|\partial_x \psi_d|^2}{2m_d} \mathrm{d}x \notag \\
			&= \frac{f_d^2 \left\{ 12 w_d^2 [1 - (x_c/R_{\rm TF})^2] -(\pi^2 + 12)/R_{\rm TF}^2 \right\}}{18 m_d w_d} -\frac{1}{m_d R_{\rm TF}} + \frac{f_d^2}{m_d w_d R_{\rm TF}^2} + \frac{1}{4m_d R_{\rm TF}} \ln \left| \frac{4 - (x_c/R_{\rm TF})^2}{(x_c/R_{\rm TF})^2} \right| \notag \\
			&-\frac{f_d^2}{2 m_d R_{\rm TF}^2} \int_{-R_{\rm TF}}^{R_{\rm TF}} \frac{\sech^2(w_d x_{th})}{1 - \left[(x_c + x_{th})/R_{\rm TF}\right]^2} {\rm d}x_{th}.
		\end{align}
		For the last term, the result diverges when directly integrated, due to the spatially varying denominator. Since the background density varies slowly over the soliton scale ($w_d \gg 1/R_{\rm TF}$), one can reasonably Taylor-expand the denominator around the soliton center $x_{th} = 0$, which yields $-\frac{f_d^2}{2 m_d R_{\rm TF}^2} \int_{-R_{\rm TF}}^{R_{\rm TF}} \frac{\sech^2(w_d x_{th})}{1 - \left[(x_c + x_{th})/R_{\rm TF}\right]^2} {\rm d}x_{th} = -\frac{f_d^2}{2 m_d R_{\rm TF}^2} \int_{-l_0}^{l_0} \{\frac{1}{1 - (x_c/R_{\rm TF})^2}[1 + C_2 x_{th}^2 + C_4 x_{th}^4 +\mathcal{O}(x_{th}^2)]\}  \sech^2(w_d x_{th}){\rm d}x_{th}$. The coefficients are $C_2 = \frac{1}{R_{\rm TF}^2\left[1 - (x_c/R_{\rm TF})^2\right]} + \frac{4x_c^2}{R_{\rm TF}^4\left[1 - (x_c/R_{\rm TF})^2\right]^2}$ and $C_4 = \frac{1}{R_{\rm TF}^4\left[1 - (x_c/R_{\rm TF})^2\right]^2} + \frac{12x_c^2}{R_{\rm TF}^6\left[1 - (x_c/R_{\rm TF})^2\right]^3} + \frac{16x_c^4}{R_{\rm TF}^8\left[1 - (x_c/R_{\rm TF})^2\right]^4}$.
		Crucially, due to the exact parity symmetry of $\operatorname{sech}^2(w_d x_{th})$, all odd-power terms identically vanish upon integration over the symmetric interval $[-l_0, l_0]$ within the Thomas-Fermi radius ($l_0 \ll R_{\rm TF}$). Furthermore, because the soliton profile decays exponentially beyond its characteristic width $1/w_d$, the strong localization condition ($w_d l_0 \gg 1$) ensures that integral contributions from $|x_{th}| > l_0$ are vanishingly small. This allows us to safely extend the integration limits to $\pm\infty$, thereby naturally removing the intermediate cutoff $l_0$. Therefore, we finally obtain the analytical result
		\begin{align}
			E_{kd} &= \frac{f_d^2 \left\{ 12 w_d^2 [1 - (x_c/R_{\rm TF})^2] -(\pi^2 + 12)/R_{\rm TF}^2 \right\}}{18 m_d w_d} -\frac{1}{m_d R_{\rm TF}} + \frac{f_d^2}{m_d w_d R_{\rm TF}^2} + \frac{1}{4m_d R_{\rm TF}} \ln \left| \frac{4 - (x_c/R_{\rm TF})^2}{(x_c/R_{\rm TF})^2} \right| \notag \\
			&-\frac{f_d^2}{2 m_d R_{\rm TF}^2} \left[ \frac{2 \left[1 + (x_c/R_{\rm TF})^2 + (x_c/R_{\rm TF})^4\right]}{w_d} + \frac{\pi^2 \left(1/R_{\rm TF}^2 + 6 x_c^2/R_{\rm TF}^4\right)}{6 w_d^3} + \frac{7 \pi^4}{120 w_d^5 R_{\rm TF}^4} \right].
		\end{align}
		Evaluating the remaining energy contributions entails direct spatial integration of the localized ansatz. The dark soliton self-interaction and potential energies are integrated as
		\begin{align}
			E_{id} &= \int_{-R_{\rm TF}}^{R_{\rm TF}} \frac{g_d}{2} \big[ |\psi_d|^2 - \left(1 - (x/R_{\rm TF})^2\right) \big]^2 \mathrm{d}x \notag \\
			&= \frac{f_d^4 g_d}{2} \left[ \frac{4 \left[1 - (x_c/R_{\rm TF})^2\right]^2}{3 w_d} + \frac{2 \left[3 (x_c/R_{\rm TF})^2 - 1\right] (\pi^2 - 6)}{9 w_d^3 R_{\rm TF}^2} + \frac{7 \pi^4 - 60 \pi^2}{180 w_d^5 R_{\rm TF}^4} \right], \\
			E_{pd} &= \int_{-R_{\rm TF}}^{R_{\rm TF}} \frac{1}{2} m_d \omega^2 x^2 |\psi_d|^2 \mathrm{d}x \notag \\
			&= -\frac{m_d \omega^2 R_{\rm TF}^3}{5} + \frac{m_d \omega^2 R_{\rm TF}^3 \left[1 - 6 (x_c/R_{\rm TF})^2\right]}{3} + m_d \omega^2 R_{\rm TF} x_c^2 \left[1 - (x_c/R_{\rm TF})^2\right] \notag \\
			&\quad + \frac{1}{2} f_d^2 m_d \omega^2 \left[ -\frac{\pi^2 + 12 w_d^2 x_c^2}{6 w_d^3} + \frac{1}{R_{\rm TF}^2} \left( \frac{7 \pi^4}{120 w_d^5} + \frac{\pi^2 x_c^2}{w_d^3} + \frac{2 x_c^4}{w_d} \right) \right].
		\end{align}
		Similarly, explicitly evaluating the characteristic localized spatial profiles, the inter-component interaction energy is rigorously reduced to:
		\begin{align}
			E_{ibd} &= \int_{-R_{\rm TF}}^{R_{\rm TF}} g_{bd} |\psi_b|^2 \big[ |\psi_d|^2 - \left(1 - (x/R_{\rm TF})^2\right) \big] \mathrm{d}x \notag \\
			&= - g_{bd} f_b^2 f_d^2 \left[1 - (x_c/R_{\rm TF})^2\right] \int_{-R_{\rm TF}}^{R_{\rm TF}} \operatorname{sech}^2(w_b x) \operatorname{sech}^2(w_d x) \mathrm{d}x \notag \\
			&\quad + \frac{g_{bd} f_b^2 f_d^2}{R_{\rm TF}^2} \int_{-R_{\rm TF}}^{R_{\rm TF}} x^2 \operatorname{sech}^2(w_b x) \operatorname{sech}^2(w_d x) \mathrm{d}x.
		\end{align}
		
		Finally, expressing the bright soliton observables in terms of its particle number $N_b = 2f_b^2 / w_b$ and soliton velocity $v = \phi$, its kinetic, self-interaction, and potential energies are straightforwardly evaluated as
		\begin{align}
			E_{kb} &= \frac{N_b}{6} (3 v^2 + w_b^2), \\
			E_{ib} &= \frac{g_b N_b^2 w_b}{6}, \\
			E_{pb} &= \frac{1}{2} N_b \omega^2 x_c^2.
		\end{align}
		
		By defining the effective particle numbers of dark soliton component $N_d = 2f_d^2/w_d$, the excitation energy $E_s = E_{s,0} + C_{s,2} x_c^2 + C_{s,4} x_c^4$, which encapsulates the kinetic energy, self-interactions, and inter-component coupling, can be expressed more intuitively as polynomials of $\omega$
		\begin{align}
			E_{s,0} &= \left[ \frac{N_d w_d^2}{3 m_d} + \frac{N_d^2 w_d g_d}{6} - \frac{N_b N_d w_b w_d g_{bd}}{4} \int_{-R_{\rm TF}}^{R_{\rm TF}} \operatorname{sech}^2(w_b x) \operatorname{sech}^2(w_d x) \mathrm{d}x + \frac{1}{6} N_b (3 v^2 + w_b^2) + \frac{1}{6} g_b N_b^2 w_b \right] \notag \\
			&\quad + \left[ \frac{1}{4 \sqrt{2 m_d g_d}} \ln \left| \frac{8 g_d - m_d \omega^2 x_c^2}{m_d \omega^2 x_c^2} \right| - \frac{1}{\sqrt{2 m_d g_d}} \right] \omega \notag \\
			&\quad + \left[ \frac{N_b N_d w_b w_d m_d g_{bd}}{8 g_d} \int_{-R_{\rm TF}}^{R_{\rm TF}} x^2 \operatorname{sech}^2(w_b x) \operatorname{sech}^2(w_d x) \mathrm{d}x - \frac{N_d (\pi^2 + 12)}{72 g_d} - \frac{N_d^2 m_d (\pi^2 - 6)}{72 w_d} \right] \omega^2 \notag \\
			&\quad + \left[ \frac{N_d^2 m_d^2 (7 \pi^4 - 60 \pi^2)}{5760 g_d w_d^3} - \frac{\pi^2 N_d m_d}{96 g_d^2 w_d^2} \right] \omega^4 - \left[ \frac{7 \pi^4 N_d m_d^2}{3840 g_d^3 w_d^4} \right] \omega^6,\\
			C_{s,2} &= \left[ \frac{N_b N_d w_b w_d m_d g_{bd}}{8 g_d} \int_{-R_{\rm TF}}^{R_{\rm TF}} \operatorname{sech}^2(w_b x) \operatorname{sech}^2(w_d x) \mathrm{d}x - \frac{N_d w_d^2}{6 g_d} - \frac{N_d^2 m_d w_d}{6} \right] \omega^2 \notag \\
			&\quad + \left[ \frac{N_d^2 m_d^2 (\pi^2 - 6)}{48 g_d w_d} - \frac{N_d m_d}{8 g_d^2} \right] \omega^4 - \left[ \frac{\pi^2 N_d m_d^2}{32 g_d^3 w_d^2} \right] \omega^6,\\
			C_{s,4} &= \left[ \frac{N_d^2 m_d^2 w_d}{24 g_d} \right] \omega^4 - \left[ \frac{N_d m_d^2}{16 g_d^3} \right] \omega^6.
		\end{align}
		The potential energy $E_p = E_{p,0} + C_{p,2} x_c^2 + C_{p,4} x_c^4$ accounts for the interaction between the dark-bright soliton complex and the harmonic trap. Substituting the effective particle numbers, the coefficients are evaluated as
		\begin{align}
			E_{p,0} &= \left( \frac{4 \sqrt{2} g_d^{3/2}}{15 \sqrt{m_d}} \right) \omega^{-1} - \left( \frac{\pi^2 N_d m_d}{24 w_d^2} \right) \omega^2 + \left( \frac{7 \pi^4 N_d m_d^2}{960 g_d w_d^4} \right) \omega^4, \\
			C_{p,2} &= - \left( \sqrt{2 m_d g_d} \right) \omega + M_p \omega^2 + \left( \frac{\pi^2 N_d m_d^2}{8 g_d w_d^2} \right) \omega^4, \\
			C_{p,4} &= - \left( \frac{m_d^{3/2}}{\sqrt{2 g_d}} \right) \omega^3 + \left( \frac{N_d m_d^2}{4 g_d} \right) \omega^4.
		\end{align}
		
		The above excitation energy of the soliton depends on both velocity and position, and the potential energy also involves moving speed indirectly. To clearly see how the position variation affects the kinetic equation for the soliton, we rewrite the total energy as
		\begin{align}
			E_{tot}(v^2, x_c)& =  E_{s,0} + C_{s,2} x_c^2 + C_{s,4} x_c^4 + E_{p,0} + C_{p,2} x_c^2 + C_{p,4} x_c^4 \notag \\
			&= E_{s,0}(v^2) + \tilde{E}_p(v^2, x_c),
		\end{align}
		where the intrinsic excitation energy $E_{s,0}(v^2)$ as a pure function of the squared velocity, and the effective potential energy $\tilde{E}_p(v^2, x_c)=E_{p,0}+ (C_{s,2}+ C_{p,2}) x_c^2 + (C_{s,4}+ C_{p,4}) x_c^4$ as a function of both the squared velocity and the spatial center. The squared velocity dependence arises from the evenness of the excitation energy with respect to velocity and the symmetry of the soliton profiles with respect to the inverse velocity.
		The energy conservation $\frac{{\rm d}E_{tot}}{{\rm d}t} = 0$ yields
		\begin{equation}
			\frac{\partial E_{s,0}}{\partial (v^2)} \frac{d(v^2)}{dt} + \frac{\partial \tilde{E}_p}{\partial (v^2)} \frac{d(v^2)}{dt} + \frac{\partial \tilde{E}_p}{\partial x_c} \frac{dx_c}{dt} = 0.
		\end{equation}
		Substituting the kinematic relations $dx_c/dt = v$ and $d(v^2)/dt = 2v\ddot{x}_c$, and factoring out the non-zero instantaneous velocity $v$, the generalized equation of motion is extracted as:
		\begin{equation}
			2 \left[ \frac{\partial E_{s,0}}{\partial (v^2)} + \frac{\partial \tilde{E}_p}{\partial (v^2)} \right] \ddot{x}_c + \frac{\partial \tilde{E}_p}{\partial x_c} = 0.
		\end{equation}
		The $\tilde{E}_p$ and $ E_{s,0}$ both generally depend on the velocity, since the soliton profiles depend on the velocity. The effective potential can be defined by $\ddot{x}_c = -\frac{\partial U_{\rm eff}(x_c)}{\partial x_c}$. The effective potential can be calculated as
		\begin{align}
			U_{\rm eff} &= \int_0^{x_c} \frac{2v(\zeta)\zeta[(C_{s,2}+ C_{p,2})+2(C_{s,4}+ C_{p,4}) \zeta^2]}{\partial_v(E_{s,0} + E_{p,0})+\partial_v(C_{s,2}+ C_{p,2})\zeta^2+\partial_v (C_{s,4}+ C_{p,4}) \zeta^4} d\zeta \notag \\
			&\quad + {\rm const.}
		\end{align}
		based on the above analyses. The trajectory of the soliton can be obtained from the general kinetic equation (see those in Fig.~1(b) of the main text). By the way, we emphasize that the general kinetic equation holds for cases where the Thomas-Fermi approximation is valid. It fails to work well when the soliton moves into the marginal region of the condensates in the harmonic trap.
		
		It is worth noting that one needs to solve for the parameters $w_{b,d}$ and  $f_{b,d}$ at different velocities in this process by employing Eqs.~\eqref{eq:reduced_dynamics_wb} and \eqref{eq:reduced_dynamics_wd} modified under the local density approximation for the inhomogeneous background. Corresponding results are the three coupled algebraic constraints
		\begin{align}
			&\sqrt{\frac{N_b w_b}{2}} \left[ f_d^2 g_{bd} \left[1 - (x_c/R_{\rm TF})^2\right] \left( G + w_b \frac{\partial G}{\partial w_b} \right) - \frac{N_b g_b}{3} - \frac{2 w_b}{3} \right] - \left[1 - (x_c/R_{\rm TF})^2\right] \left( f_d \sqrt{1 - f_d^2} - \arcsin f_d \right) \dot{x}_c = 0, \label{eq:lda_constraint_fb}\\
			&- \frac{1}{2} N_b w_b g_{bd} G + \frac{4 f_d^2 g_d \left[1 - (x_c/R_{\rm TF})^2\right]}{3 w_d} + \frac{2 w_d}{3 m_d} - \frac{\sqrt{2 N_b w_b} f_d \dot{x}_c}{\sqrt{1 - f_d^2}} = 0, \label{eq:lda_constraint_fd}\\
			&\frac{2}{3 m_d} - \frac{2 f_d^2 g_d \left[1 - (x_c/R_{\rm TF})^2\right]}{3 w_d^2} - \frac{1}{2} N_b w_b g_{bd} \frac{\partial G}{\partial w_d} = 0. \label{eq:lda_constraint_wd}
		\end{align}
		By self-consistently solving these transformed equations for a specified set of interaction strengths and the instantaneous velocity $v = \dot{x}_c$,   all the expansion coefficients can be determined.
	\end{appendix}
\end{widetext}

\end{document}